\documentclass[]{aastex631}

\newcommand{\GJtwelve}{GJ~1214~b}
\newcommand{\mum}{\mu{\rm m}}
\newcommand{\Teq}{{\rm T}_{\rm eq}}

\shorttitle{Hazes as a probe of C/O ratio}
\shortauthors{Corrales \& Gavilan et al.}

\begin{document}

\title{Photochemical hazes can trace the C/O ratio in exoplanet atmospheres}

\correspondingauthor{L.~Corrales}
\email{liac@umich.edu}

\author[0000-0002-5466-3817]{L\'ia Corrales}
\affil{University of Michigan, Dept. of Astronomy, 1085 S University Ave, Ann Arbor, MI 48109, USA}

\author[0000-0001-8645-8415]{Lisseth Gavilan}
\affiliation{NASA Ames Research Center, Space Science \& Astrobiology Division, Moffett Field, CA 94035, USA}
\affiliation{Bay Area Environmental Research Institute (BAERI), Sonoma, CA 95476, USA}

\author[0000-0002-1912-3057]{D. J. Teal}
\affil{University of Maryland, Department of Astronomy, 4296 Stadium Dr, College Park, MD 20742, USA}

\author[0000-0002-1337-9051]{Eliza M.-R. Kempton}
\affil{University of Maryland, Department of Astronomy, 4296 Stadium Dr, College Park, MD 20742, USA}

\begin{abstract}

Photochemical hazes are suspected to obscure molecular features, such as water, from detection in the transmission spectra of exoplanets with atmospheric temperatures $< 800$~K. 
The opacities of laboratory produced organic compounds (tholins) from \citet{Khare1984} have become a standard for modeling haze in exoplanet atmospheres. However, these tholins were grown in an oxygen-free, Titan-like environment that is very different from typical assumptions for exoplanets, where C/O$\sim 0.5$. 
This work presents the $0.13-10~\mum$ complex refractive indices derived from laboratory transmission measurements of tholins grown in environments with different oxygen abundances. With the increasing uptake of oxygen, absorption increases across the entire wavelength range, and a scattering feature around $6~\mum$ shifts towards shorter wavelengths and becomes more peaked around $5.8~\mum$, due to a C=O stretch resonance. Using \GJtwelve\ as a test-case, we examine the transmission spectra of a sub-Neptune planet with C/O ratios of solar, 1, and 1000 to evaluate the effective differences between our opacities and those of Khare. 
For an atmosphere with solar hydrogen and helium abundances, we find a difference of 200-1500~ppm, but for high-metallicity (Z=1000) environments, the difference may only be 20~ppm. The $1-2~\mum$ transmission data for \GJtwelve\ rule out the Titan-like haze model, and are more consistent with C/O$=1$ and C/O$=$solar haze models. This work demonstrates that using haze opacities that are more consistent with underlying assumptions about bulk atmospheric composition are important for building self-consistent models that appropriately constrain the atmospheric C/O ratio, even when molecular features are obscured.

\end{abstract}

\keywords{Exoplanet atmospheric composition (487) --- Laboratory astrophysics (2004) --- Astrochemistry (75) --- Exoplanet evolution (491)}

\section{Introduction} \label{sec:intro}

Approximately 75\% of the over 5,000 exoplanets known today were discovered via the transit method, where the chance alignment of an extra-solar planet's orbit with Earth's vantage point causes the planet to pass in front of the star, blocking $\sim 0.1-1$\% of its light. 
Observing a transit at multiple wavelengths builds a transmission spectrum, on which the contents of the exoplanet's atmosphere are imprinted via their unique spectroscopic fingerprints in absorption and scattering. 
The depth of an exoplanet transit, as a function of wavelength, depends jointly on the transmission properties of the atmospheric contents as well as their vertical distribution.
In this letter, we investigate whether the spectral features of atmospheric hazes on an exoplanet can provide key markers of the bulk content of the gas in which they form.

Aerosols -- whether condensing directly from atmospheric gas (clouds) or through photochemical reactions (hazes) -- are known to affect nearly every type of exoplanet atmosphere. 
Even hot Jupiters, across a wide range of temperatures ($1000 - 2000$~K), have spectral features that are muted as a result of aerosol obscuration \citep{Sing2016} 
and steep optical slopes that are suspected to arise from a combination of clouds and hazes \citep{Sing2011, Pont2013, McCullough2014, SanchezLopez2020, Steinrueck2021}. 
Theoretical models predict that the infrared opacity of hot Jupiters with $\Teq \sim 900-2200$~K will be dominated by mineral condensates rich in refractory elements \citep{Helling2019, Helling2016, Powell2018, Gao2020}. Only below about 800~K are photochemically produced organic hazes expected to form and dominate the infrared opacity \citep{Morley2015, Gao2020}.

Photochemical hazes are also suspected to be a key source of opacity for more temperate, smaller planets such as sub-Neptunes and super-Earths. 
One such planet is \GJtwelve, which has a mass of 6.55~M$_{\rm E}$ and radius of 2.68~R$_{\rm E}$, consistent with a variety of composition models that suggest it hosts an atmosphere comprising $\sim 0.5$ to a few percent of the planet's mass \citep{Charbonneau2009, Rogers2010}. 
Transmission measurements from $0.7-4.5~\mum$ can be reproduced by models employing some combination of a high mean molecular weight atmosphere and an optically thick aerosol layer at altitudes $\sim 10$~mbar \citep{Bean2010, Desert2011, Kempton2012, Fraine2013, Morley2013, Kreidberg2014}.
The chemical composition and origin of the obscuration is not well-known, and models of cloud condensation in exoplanet atmospheres can require strong loft and low sedimentation efficiency to reproduce the flat spectrum of \GJtwelve\ \citep{Morley2013, Gao2018, Ohno2018}. 
Since hazes are produced photochemically at higher altitudes, they are also under investigation to explain the flat $1-2~\mum$ transmission through the atmosphere of this planet \citep{Morley2015, Kawashima2018, Kawashima2019b, Lavvas2019, Ohno2019}.

We examine the ability of hazes to mute molecular signals and contribute their own features to the mid-IR transmission spectra of warm exoplanet atmospheres. 
As of now, the dominant opacities used to incorporate the transmission effects of photochemical hazes in models of exoplanet atmospheres mainly come from \citet[][hereafter referred to as K84]{Khare1984}, which was obtained from laboratory grown compounds (tholins) in a simulated Titan atmosphere -- a majority N$_2$ environment with trace CH$_4$. The broad wavelength range covered by the K84 model has made it particularly useful for the exoplanet community, but it has fundamental limitations. 
K84 tholins exhibit a clear transmission window around $0.5-3~\mum$, but in-situe measurements of Titan hazes show more uniform absorption across $0.5-1.5~\mum$ \citep{Brasse2015}, in agreement with more recent laboratory measurements \citep{Tran2003b, Lavvas2010, Rannou2010, Gavilan2018}. 
This demonstrates the need for a larger variety of lab-measured aerosol optical properties, which are important for planning and interpreting observations of exoplanet atmospheres.

In this work, we showcase the attenuation properties of tholins grown in different mixtures of N$_2$, CO$_2$, and CH$_4$, providing benchmark spectral features of hazes from a variety of oxidation states. 
We apply the optical properties derived from this work to simulate the $0.3-10~\mum$ transmission spectrum of sub-Neptune \GJtwelve\ under different C/O and H+He abundance fractions to identify spectral features from hazes that provide markers for the C/O ratio of the atmosphere. 
The C/O ratio is a key tracer of atmospheric composition and can also be an indicator of where the planet formed in the protoplanetary disk, and whether its atmosphere is primordial or secondary \citep[e.g.,][]{Oberg2021}. 
With this work, we are releasing the lab-measured optical constants and attenuation cross-sections for tholins produced at three C/O ratios, which are of broad relevance to the exoplanet community.

\section{Overview of Laboratory Measurements}
\label{sec:lab}

Computing the attenuation properties of aerosols  
requires knowledge of the substance's dielectric properties, which are conveniently encoded by the real and imaginary parts of the complex index of refraction: $n^* (\lambda) = n(\lambda) + i k(\lambda)$. In applying $n^*$ to the wave equations for light propagation through a medium, the imaginary $k$ component causes the electric field to decay exponentially with distance (absorption) and the real part $n$ induces a phase shift (scattering). Throughout this section, we compare the $n$ and $k$ spectrum of tholins as proxies for the significance of scattering and absorption, respectively. 
Generally, experiments that form tholins with CO or CO$_2$ agree on the overall impact of increasing Oxygen: the real optical index $n$ increases towards shorter wavelengths, while $k$ makes them more absorbing in the UV-Vis \citep{Hasenkopf2010, Ugelow2018, Gavilan2017, Gavilan2018, Jovanovic2021}.

\citet[][G17]{Gavilan2017} and \citet[][G18 hereafter]{Gavilan2018} investigated the role of atmospheric CO$_2$ on the optical properties of tholins 
prepared using the PAMPRE chamber \citep{Szopa2006} located at LATMOS (U. Paris-Saclay, France). In this chamber, the neutral gas remains at room temperature ($\sim$300 K) while the electrons have a mean energy of 1-2 eV \cite[$\sim 10^4$~K,][]{Alcouffe2010}. 
These temperatures span the range of estimated atmospheric temperature profiles for  \GJtwelve\ \citep{Kempton2010,Kawashima2019b}.  
For these experiments, an increasingly oxygenated atmosphere was created by increasing the CO$_2$/CH$_4$ ratio from 0 to 4, while keeping a constant molar fraction of N$_2$. 
G17 used the ellipsometry technique to measure both the $n$ and $k$ values in the $270-600$~nm wavelength range.
Through UV-MIR transmission spectroscopy, G18 obtained a direct measurement of the $k$ values from the broader wavelength range of 130~nm to 10~$\mum$. 
This latter study revealed absorption resonances spanning the vacuum-ultraviolet (VUV) to the mid-infrared (MIR). Electronic transitions in the $200-500$~nm range were attributed to amine groups and, as the CO$_2$/CH$_4$ ratio increases, to electronic transitions from hydroxyl (-OH) and carboxyl (-COOH) groups. For the most oxygen-rich samples, absorption is greatest in the $0.13-0.3~\mum$ and $6-10~\mum$ regions. 

\subsection{Derivation of optical constants}
\label{sec:optical-constants}

We present the complex refractive indices of three tholin samples from G18: those produced in an N$_2$:CO$_2$:CH$_4$ mixture of 95:0:5 (C/O=$\infty$), 90:5:5 (C/O$=1$), and 90:8:2 (C/O$=0.625$, which is near-solar). 
The imaginary part of the complex of index of refraction was derived from transmission measurements obtained in four wavelength ranges: 
the vacuum-ultraviolet to UV ($130-250$~nm), the UV-Vis ($210-1000$~nm), the near-infrared ($1.05-2.7~\mum$), and the mid-infrared ($1.43 - 10~\mum$). For the region with no data ($1-1.05~\mum$) we interpolated between the visible and near-IR data. Due to the different spectral resolution of each wavelength range, data was interpolated onto a new regularized grid of 1000 wavenumber values, logarithmically spaced from $0.13 - 10~\mum$. 

The final composite $k$ spectrum was used to calculate the $n$ spectrum, via the Kramers-Kronig 
relations \citep{Kronig1926, Keefe2002, Lucarini2005}.
We use
OpC\footnote{\url{https://github.com/zmeri/opC}}, which is 
based on the FORTRAN program LZKKTB \citep[also known as KKTRANS,][]{Bertie1992}, and is modified to include the non-constant electronic contribution to the real refractive index discussed in \citep{Bertie1995}.
As part of the OpC calculation, the $k$ spectrum is linearly extrapolated down to 0 at the wavenumber of 0 \citep{Bertie1992}. 
It uses the Maclaurin method to numerically calculate the Cauchy principal value of the integral which improves the accuracy of the transform near intense absorption peaks \cite{Ohta1988}. The transform requires an ``anchor'' value for the real part of the complex index of refraction at high wavenumber. Because we lack a precise measurement of $n(\lambda > 10~\mum$), we use $n(600{\rm nm})$ calculated from the ellipsometry experiment in G17. 
The direct transmission measurements are considered highly reliable, and the $n$ values scale linearly with the choice of the anchor value, so we estimate an uncertainty of $\pm 2.5\%$ on $k$ and $\pm 5\%$ on most of the $n$ values. The uncertainty on $n$ is likely higher at the endpoints of the wavelength range, $\pm 15\%$, due to the extrapolations employed by OpC.

\begin{figure*}
    \centering
    \includegraphics[width=\textwidth]{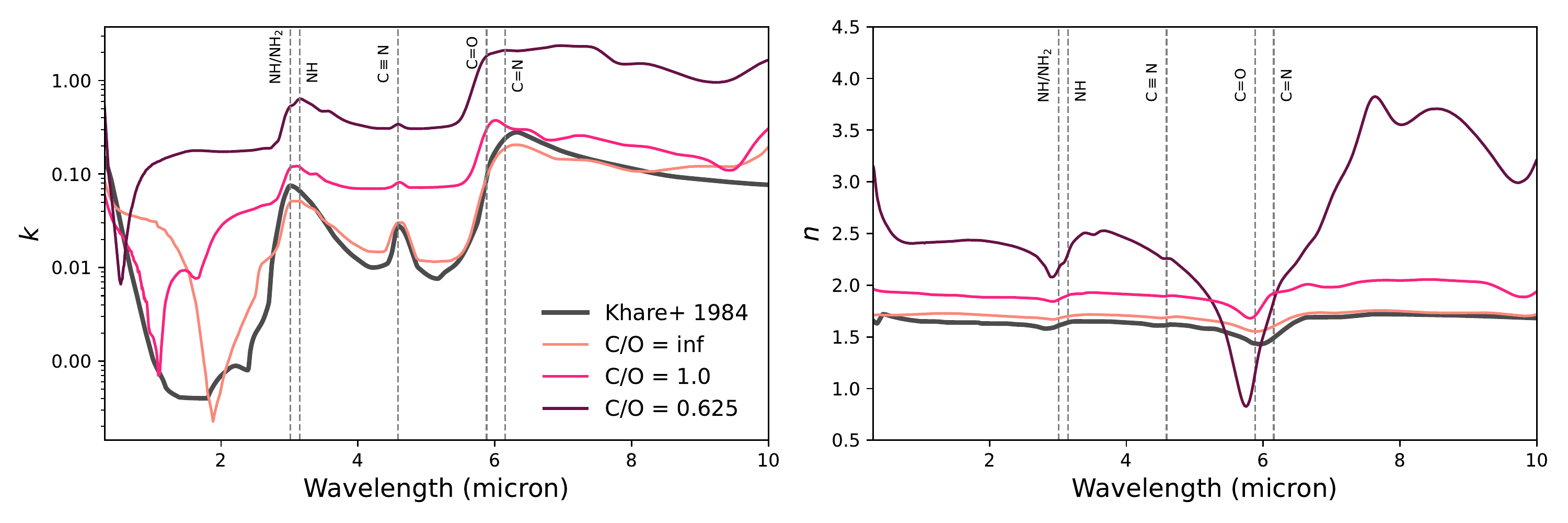}
    \caption{
    The imaginary ($k$, left) and real ($n$, right) parts of the complex index of refraction as measured for hazes produced from laboratory gas mixtures with different C/O ratios (G18). The optical constants derived from the laboratory work of \citep{Khare1984} are overlaid for reference. The resonant wavelength for various molecular stretching and vibrational bands, suspected to underlie the main infrared absorption features, are identified with dash vertical lines in each plot. 
    }
    \label{fig:complex_index}
\end{figure*}

Figure~\ref{fig:complex_index} shows the results of the OpC calculation for the real ($n$) part of the complex index of refraction, given the imaginary part ($k$) measured in the lab. 
To remove a few zero values in the $k$ curve, we smoothed all optical constants using the Savitzky-Golay algorithm, employing a fourth order polynomial least-squares fit across 11 adjacent bins at every data point \citep{SGsmooth, NumRecipes}. 
To mimic the growth of hazes in the oxygen-free (C/O$=\infty$) environment of Titan, K84 used slightly different gas abundances, N$_2$:CH$_4$ = 90:10 (K84) versus 95:5 (G18). 
Nonetheless, the optical constants of the hazes produced in a  C/O=$\infty$ environment by G18 are within agreement with K84 at a level that is consistent with the variations found throughout the literature and within the environment of Titan, itself \citep{Brasse2015,Lavvas2010, Rannou2010}. 

Figure~\ref{fig:complex_index} also identifies some of the major mid-infrared vibrational absorption bands.
The intensities and positions of the vibrational bands observed from the hazes change as CO$_2$ is added to the haze-growing environment. As the oxygen content increases in the laboratory environment, the oxygen content of the hazes also appears to increase, as evidenced by the strong C=O features. Meanwhile, the overall contrast of the NH and C=N features around $3~\mum$ and $4.6~\mum$, respectively, becomes less prevalent when the hazes are grown in a more oxygen rich environment. 
Adding a moderate amount of CO$_2$ (C/O$ = 1$) causes a shift the of the mid-infrared absorption peak towards a C=O stretching mode at $5.88~\mum$. Adding even more CO$_2$ (C/O$ = 0.625$) greatly enhances the haze absorptivity across all wavelengths considered. For this near-Solar C/O environment, a variety of stretching and bending modes from C=N, C=O, and C=C overlap, resulting in relatively flat continuum absorption for wavelengths longer than $6~\mum$. This feature of the spectrum creates strong scattering resonances near $6~\mum$, due to anomalous dispersion. 
For a more complete identification and comparison of the spectral features found in the tholins shown here, we refer the reader to the original paper by G18.

\subsection{Calculation of particle cross-sections}
\label{sec:cross-sections}

The $n$ and $k$ values calculated in Section~\ref{sec:optical-constants} are used as inputs for calculating the absorption and scattering cross-sections. We employ the \citet{BHMie} algorithm for the general Mie solution for computing the absorption and scattering of spherical particles, using the \texttt{newdust} Python library for generic multi-wavelength extinction by astrophysical particulates \citep{Corrales2016, eblur-newdust}\footnote{This code employs a vector-based computation of the original \citet{BHMie}  algorithm. It is open source and publicly available at https://github.com/eblur/newdust} 
We find that scattering is generally negligible for the small particles around $1-10$~nm, making it so that their extinction cross-sections roughly follow the absorption profile exactly, displaying all the features of Figure~\ref{fig:complex_index}. For larger $\sim 0.1~\mum$ particles, scattering dominates over absorption at wavelengths shorter than $2~\mum$, leading to roughly featureless transmission. However, extinction features from 3.4~$\mu{\rm m}$ to 6~$\mu{\rm m}$ may still be used to identify haze species from these larger particles.

We calculated the attenuation efficiency, which relates the cross-section for a physical interaction to the projected geometric cross-section of the particle ($Q = \sigma / \pi a^2$, where $a$ is the particle radius), for a range of particle sizes between $1$~nm and $10~\mum$ over the wavelength range of 130~nm to 10~$\mum$. The scattering, absorption, and total extinction (absorption plus scattering) efficiencies are publicly available in ASCII and FITS file format.\footnote{https://doi.org/10.5281/zenodo.7500026} This archive also provides the geometric scattering factor, $g = \langle \cos \theta \rangle$, for each particle size and wavelength. Pure forward scattering is characterized by $g=1$ and isotropic 
scattering is characterized by $g=0$. The $g$ value is relevant for deciding how much light is effectively removed from the path of incident radiation, which determines whether or not scattering contributes to the effective opacity of a medium.

A value of $g \geq 0.8$ could lead to a $10\%$ difference in the transmission properties for some hot Jupiter or sub-Neptune sized planets that are accessible for transit measurements today \citep{Roberts2017}. We find that this condition is mainly satisfied for particles $> 1~\mum$ at wavelengths $< 500$~nm. 
In Section~\ref{sec:simulations} of this work, we use the vertical haze particle distributions of \citep{Kawashima2019b}, computed for \GJtwelve, and simulate its transmission properties from 300~nm to 10~$\mum$ using a version of ExoTransmit that is modified for aerosols \citep{Kempton2017, Teal2022}. We find that the majority of haze particles for these simulations have radii $< 1~\mum$ for the region of the atmosphere that is not optically thick ($P < 1$~mbar) at short wavelength. Furthermore, the focus of this work is to identify NIR-MIR spectroscopic features of aerosols that could provide a marker of the atmosphere's C/O ratio. 
For all these reasons, we use the total extinction cross-section ($Q_{\rm abs} + Q_{\rm sca}$) to compute the transmission properties for \GJtwelve.

\section{Transmission properties of \GJtwelve\ with different C/O ratios}
\label{sec:simulations}

\begin{table}
\centering
\caption{ExoTransmit calculation inputs}
\label{tab:ModelParameters}
	\begin{tabular}{| l | c c | c | c |}
	\hline
	 This work & \multicolumn{2}{c |}{Chemical Profile} & Haze Profile & Haze Opacities \\
	\hline
	{\bf Name} & {\bf C/O}	& {\bf Z (solar)}		& {\bf KI19 model} 	& {\bf G18 setup} \\
	\hline
	Solar			& 0.5			& 1				& Fiducial			& C/O = 0.625	\\
	High-Z		& 0.5			& 1000			& 1000 $\times$ solar	& C/O = 0.625 \\
	C/O = 1		& 1			& 1				& C/O = 1			& C/O = 1 \\
	C/O = 1000	& 1000		& 1				& C/O = 1000		& C/O = $\infty$ \\
	\hline
	\end{tabular}
\end{table}

We model the transmission of \GJtwelve's atmosphere using a modified version of ExoTransmit \citep{Kempton2017} that incorporates the vertical profile of a single haze species, given number density and particle radius as a function of pressure in the atmosphere \citep{Teal2022}. 
The background atmosphere is composed of gas in thermochemical equilibrium.  The volume mixing ratios of the gas species are computed using the chemical equilibrium code of \citet{mbarek16}, given a set of elemental abundances defined by the metallicity and C/O ratio of the atmosphere.
The atmosphere is modeled with an isothermal temperature of 500~K across the pressure range of $100$ to $10^{-9}$~bar.  Gas phase absorption is calculated using the default opacity tables provided with ExoTransmit. Table~\ref{tab:ModelParameters} describes the input parameters for each set of models, defined by the metallicity, C/O ratio, and haze inputs. We use the vertical haze profiles computed by \citep[][hereafter referred to as KI19]{Kawashima2019b}, which examined the growth of hazes in a simulated \GJtwelve\ atmosphere under the influence of a variety of C/O ratios and metallicities (their Table~1). For simulations that utilized non-solar C/O ratios, the remaining metal abundances were set to their solar values relative to hydrogen. For all cases, we compare the modeled transmission under the effect of no haze, K84 haze, and the G18 haze described in Table~\ref{tab:ModelParameters}.
\begin{figure}
    \centering
    \includegraphics[]{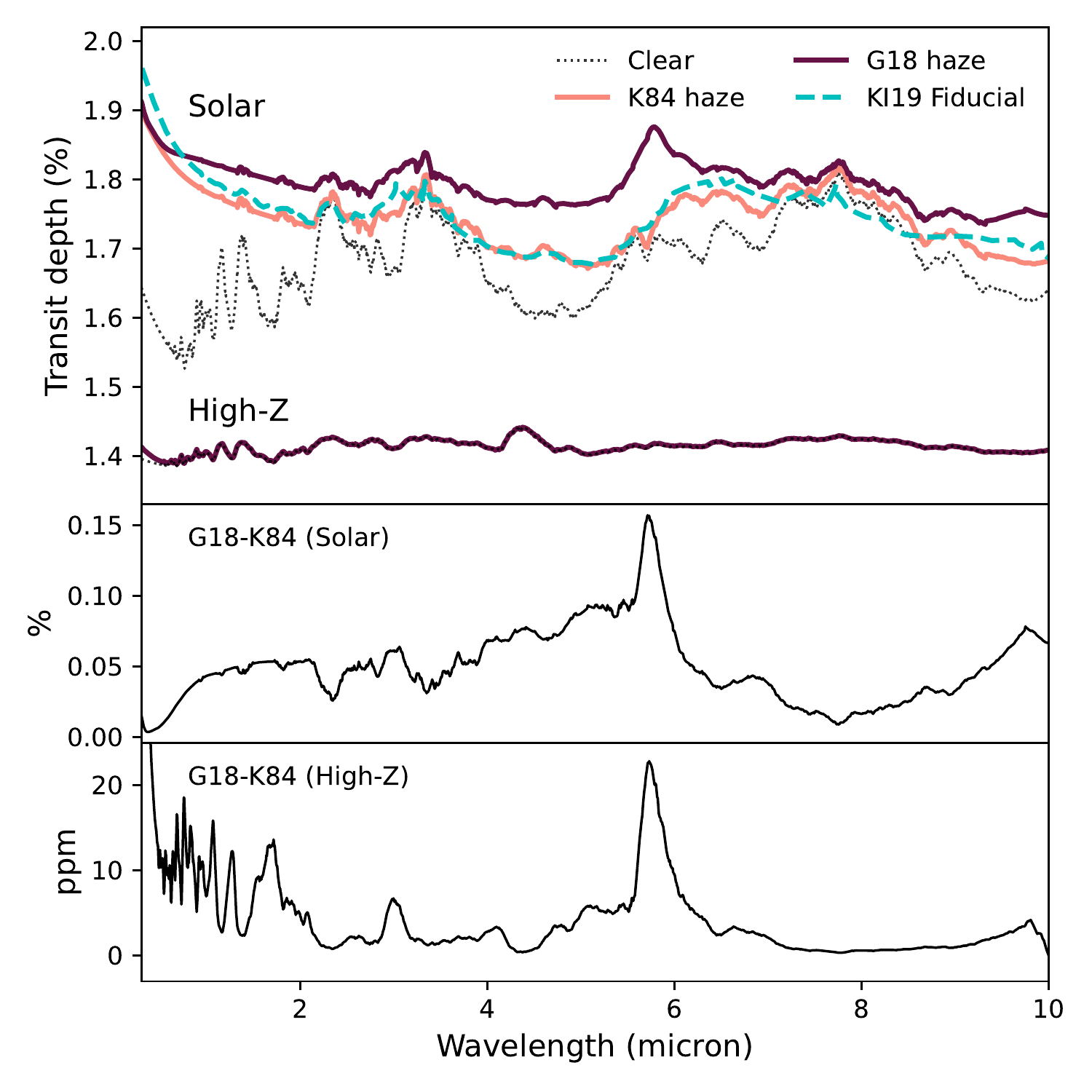}
    \caption{
    Transmission spectra of \GJtwelve\ for the fiducial case of a solar C/O ratio. 
    {\sl Top:} The transmission spectrum of \GJtwelve\ was computed under the assumption of solar metallicity relative to hydrogen (Solar, top curves) and for Z=1000 $\times$ solar metallicity (High-Z, bottom curves). The transmission spectrum produced by KI19 for their fiducial haze case, which utilized K84 opacities, is shown for comparison (dash cyan curve). This curve was scaled up by a factor of 20\% to match the K84 haze spectrum around $4.3~\mum$, which accounts for slight differences in the assumed planet parameters.
    {\sl Middle:} Using the optical constants of lab-grown tholins \citep[][G18]{Gavilan2018} with a C/O ratio close to solar leads to significantly enhanced attenuation, increasing the transit depth of \GJtwelve\ by $0.5-0.15\%$ across the wavelength range of $1-6~\mum$, relative to models computed with the optical constants of \citep[][K84]{Khare1984}. 
    {\sl Bottom:} In the High-Z atmosphere, which is highly depleted of Hydrogen and Helium, the differences between using the G18 and K84 optical properties are more subtle. Using the optical properties of G18 tholins leads to a $10-20$~ppm difference in the predicted transit depth. 
    }
    \label{fig:COsolar}
\end{figure}

There is a fundamental limitation in the sub-Neptune model assumptions that make it difficult to provide physical consistency between the molecular abundances and haze composition in the simulated spectra. None of the chemical profiles used in this work provided the 90\% N$_2$ atmospheric environment used to grow tholins. However, our goal is to examine how transmission features might change as a result of increasing oxygen uptake by the hazes, making the relative abundances of CO$_2$ and CH$_4$ of particular interest. We examined the vertical profiles of the CO$_2$/CH$_4$ ratio from both our chemical equilibrium models and the models of KI19 to see how they compared with the molecular abundances used by G18. In the Solar and C/O=1 models, CO$_2$/CH$_4 \approx 4$ and 1, respectively, at pressures around $10^{-7}$~bar, where haze particles begin to form. In the High-Z  model, CO$_2$/CH$_4 \approx 4$ at $10^{-7}$~bar and deeper, maintaining the appropriate ratio where hazes form and continue to grow. We examined the vertical profiles from a contrived mixture of C:N:O=10:180:16, designed to mimic the bulk abundances from the G18 experiment with C/O=0.625. In this case, CO$_2$/CH$_4 < 10^{-5}$ across all pressure scales. 
Since we are unable to produce a model atmosphere of \GJtwelve\ that is identical to the laboratory setup, which would also be difficult to compare with KI19, we opt to use the chemical profiles built from solar C/N abundances, which yield  CO$_2$/CH$_4$ ratios that are closer to those used in the laboratory environment.

\begin{figure}
    \centering
    \includegraphics[]{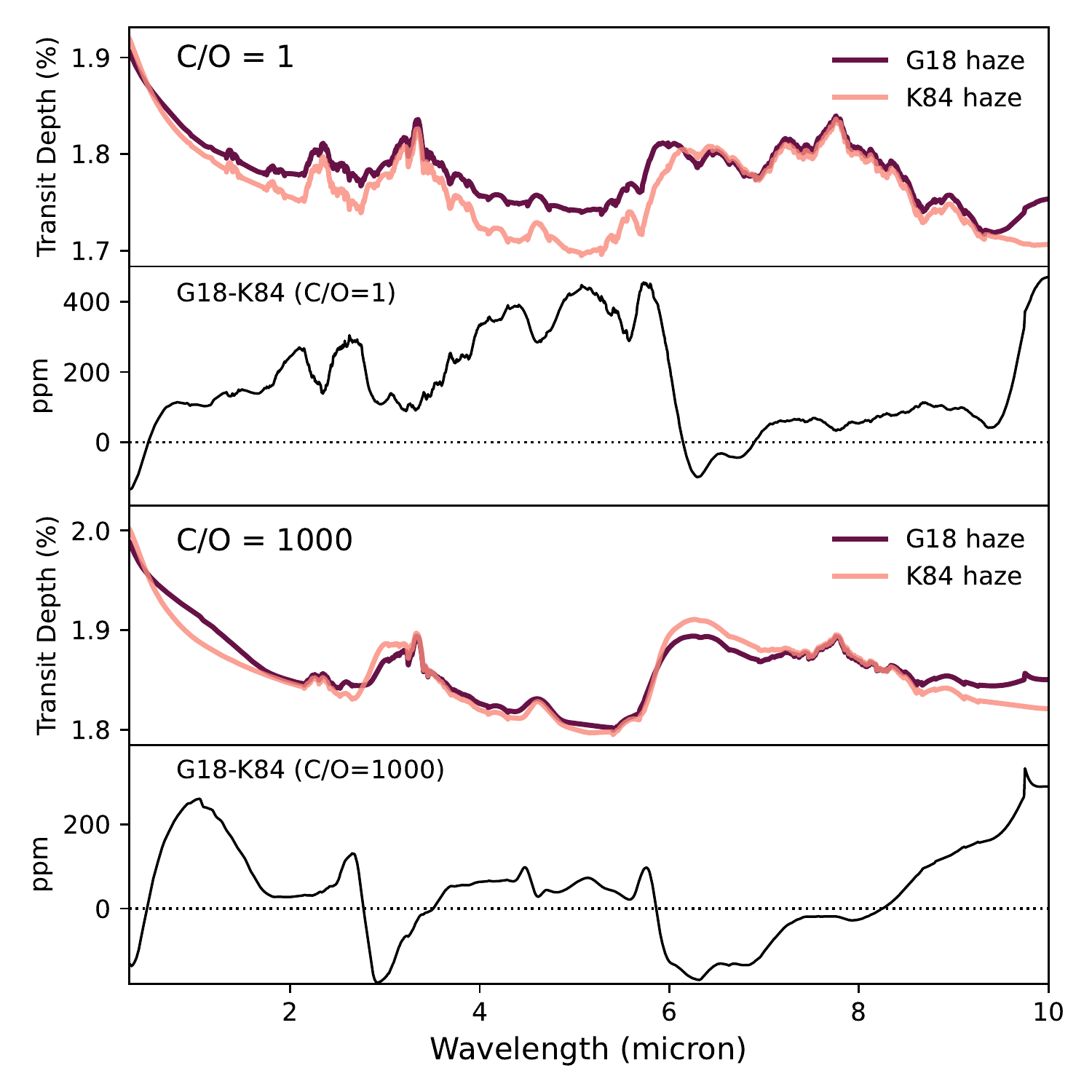}
    \caption{
    Transmission spectra of \GJtwelve\ for non-Solar C/O ratios.  
    {\sl Top two panels:} The transmission spectrum of \GJtwelve\  computed under the assumption of solar metallicity relative to hydrogen and C/O$=1$, using the K84 and G18 tholin opacities. Using the optical constants from tholins grown in a C/O$=1$ environment suggest a deeper transit than expected when using K84 opacities, especially across $2-6~\mum$, by about $200-400$~ppm.  
    {\sl Bottom two panels:} The transmission spectrum of \GJtwelve\  computed under the assumption of solar metallicity relative to hydrogen and C/O$=1000$, using the K84 and G18 tholin opacities. Since both sets of tholins were grown in a Titan-like C/O$=\infty$ environment, the two curves agree within 200~ppm. 
    }
    \label{fig:COalt}
\end{figure}

Figures~\ref{fig:COsolar}--\ref{fig:COalt} showcase the ExoTransmit results. Even with hazes, some molecular line features present themselves when modeled with the highest spectral resolution possible ($R = 1000$ for the default ExoTransmit opacity tables). For ease of visual comparison between this work and KI19, each spectrum has been smoothed via the Savitzky-Golay algorithm to remove the high resolution line features. 

Figure~\ref{fig:COsolar}  demonstrates that, despite many differences in the model complexities implemented by KI19 and this work, we were able to reproduce the transmission spectrum from the KI19 fiducial model (dash cyan curve), utilizing K84 haze opacities (peach curve), to sufficient agreement.\footnote{The normalization of the KI19 model was adjusted by 20\% to agree around $4.3~\mum$. This adjustment is necessary to account for minor differences between this work and KI19 in the assumed radius, mass, and temperature profile for \GJtwelve.} 
KI19 did not implement an isothermal temperature profile and included the effects of photochemistry, making it so that the majority of molecular species were dissociated above pressures of $10^{-7}$~bar. We tested the impact of this difference by computing the ExoTransmit spectrum with a pressure cut-off of $10^{-7}$~bar, and found no appreciable difference. 
KI19 also assume a different set of haze precursor molecules -- HCN, C$_2$H$_2$, and CH$_4$ -- than the G18 experiments, which utilized N$_2$, CH$_4$, and CO$_2$. This leads the KI19 vertical profiles of volume mixing ratios for molecules like HCN and C$_2$H$_2$, in comparison to our chemical equilibrium models, to differ by orders of magnitude, where KI19 abundances were generally higher. This might explain the differences between the transmission spectrum continuum in Figure~\ref{fig:COsolar}. 
Fortunately, the $\sim 300$~ppm differences between the KI19 fiducial and Solar (K84 haze) transmission spectrum are not a subject of concern for this work, which seeks only to compare the results of modeling transmission with different haze species. 
Haze is the leading order effect in shaping 
the transmission spectra computed in this work. 

Figure~\ref{fig:COsolar} showcases the transmission spectrum results for the Solar and High-Z models described in Table~\ref{tab:ModelParameters}. 
The transmission spectrum computed with the optical constants from the G18 tholins produced in a near-solar C/O ratio environment is significantly higher, flatter, and contains different spectral features from the transmission model that uses K84 tholin properties. In particular, the absorption and scattering resonances induced by abundant C=O bonds produce, in effect, a transit depth feature that is enhanced by $0.15\%$ around $5.8~\mum$. Continuum absorption also enhances the transit depth by about $0.05-0.10\%$ across  $1-5~\mum$ and by $0.05\%$ around $9-10~\mum$. It's particularly interesting that the G18 tholin opacities produce flatter transmission spectra overall, which may help to interpret the stronlgy featureless observed transmission spectrum of \GJtwelve\ (Section~\ref{sec:conclusions}).

Our fiducial Solar model assumes that \GJtwelve\ has a substantial H and He envelope. If \GJtwelve\ is depleted of H and He (the High-Z model), no matter what haze species is implemented, the transmission spectrum is relatively flat and featureless due to the high mean molecular weight of the atmosphere. The bottom panel of Figure~\ref{fig:COsolar} shows that it would require 20~ppm level precision to distinguish between haze species of different C/O ratios, using the $5.8~\mum$ C=O resonance feature.

Figure~\ref{fig:COalt} demonstrates that, as the elemental abundance of oxygen falls, the differences between the optical constants derived from G18 and K84 tholins become less dramatic (also seen in Figure~\ref{fig:complex_index}). In the case of C/O$=1$, using  G18 tholins in the model predict a transit depth that is enhanced by up to 400~ppm (0.04\%) across the $1-6~\mum$ range and again at $10~\mum$, compared to K84. 
Even though both K84 and G18 optical properties were determined from tholins grown in an environment free of oxygen, we used the chemical and haze vertical profiles computed with C/O$=1000$ in order to match the KI19 models. We find that the optical properties derived from G18 tholins still differ from the results obtained with K84 opacities, producing $\sim \pm 250$~ppm differences in the modeled transmission depths (Figure~\ref{fig:COalt}).

\section{Conclusions}
\label{sec:conclusions}

This work examines how spectral signatures of photochemical hazes, as informed by laboratory measurements, might constrain the C/O ratio of the atmosphere in which they formed.
The pioneering work of \citet{Khare1984} provided broad-band optical properties for tholins, grown in an oxygen-free environment, and is a dominant template for photochemical haze widely in use for exoplanet transmission models today. Employing these optical properties in models of exoplanet atmospheric transmission comes with a biased assumption that those atmospheres are nearly devoid of oxygen. 
In Section~\ref{sec:lab}, we present the $0.3-10~\mum$ optical properties of tholins grown in a gas chamber with increasing amounts of oxygen \citep{Gavilan2017, Gavilan2018}. 
These tholins exhibit infrared spectral shapes around $1-3~\mum$, $6~\mum$, and $10~\mum$ that are distinct from each other and from K84.

\begin{figure}
    \centering
    \includegraphics[width=0.6\textwidth]{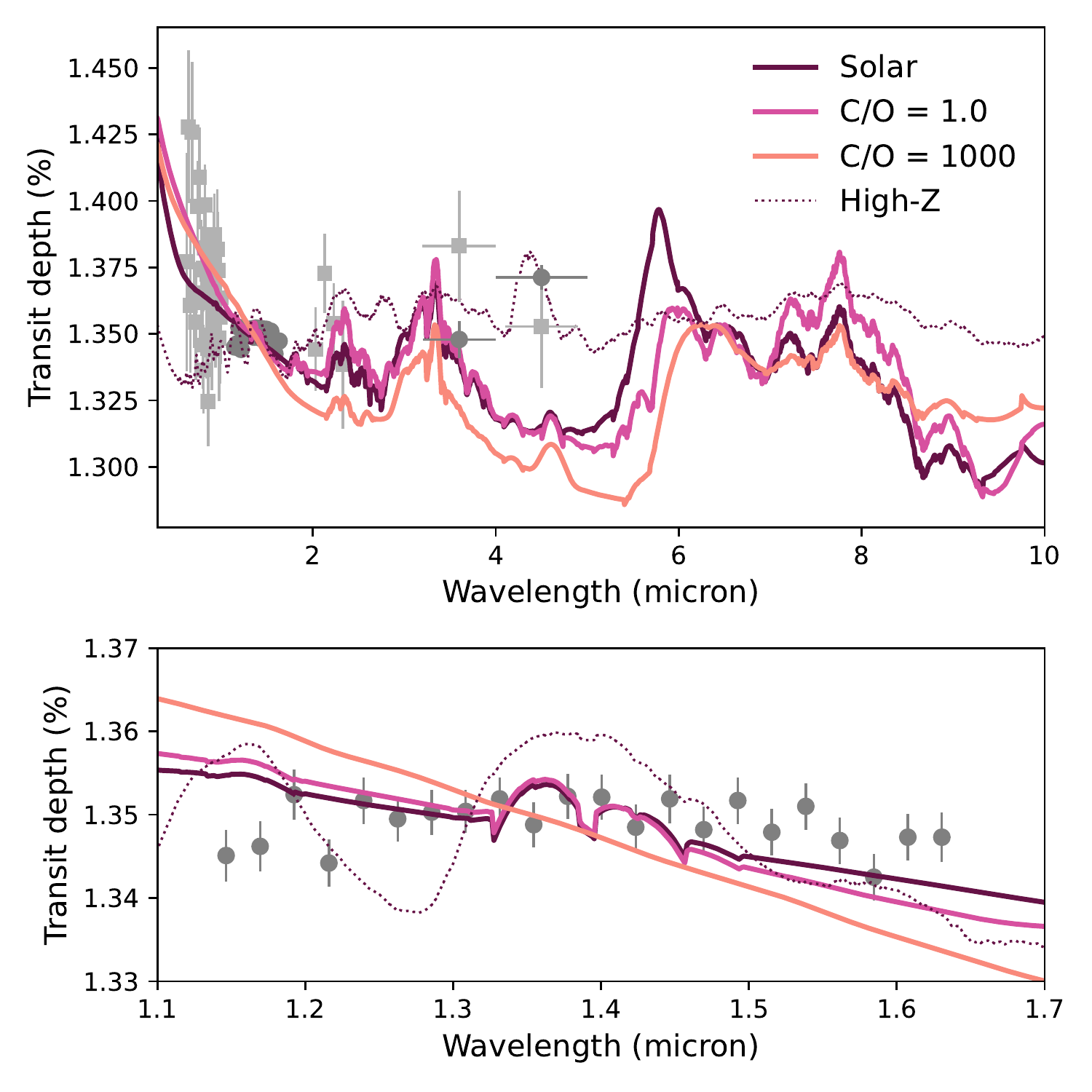}
    \caption{The transmission spectrum of \GJtwelve, with light grey squares \citep{Bean2010, Bean2011, Desert2011} and dark grey circles \citep{Fraine2013, Kreidberg2014}, overlaid with the three ExoTransmit models utilizing the new optical constants derived from the laboratory measurements of G18.
    Each ExoTransmit model has been renormalized to match the mean $1-2~\mum$ transit depth so that the shape of the spectral features can be compared. The more precise $1-2~\mum$ data are consistent with C/O ratios of Solar and 1 assuming significant Solar H/He abundances (bottom panel), but the Spitzer data points are more consistent with the $Z=1000$ model.}
    \label{fig:GJ1214}
\end{figure}

In Section~\ref{sec:simulations}, we employ the new lab-measured tholin opacities to model the transmission spectrum of sub-Neptune \GJtwelve, finding that these haze species are distinguishable from the K84 model by 200-1500~ppm, assuming Solar abundances of H and He. 
Figure~\ref{fig:GJ1214} shows the observed optical-IR transmission spectrum of \GJtwelve\ \citep{Bean2010, Bean2011, Desert2011, Fraine2013, Kreidberg2014} with the three ExoTransmit models of different C/O ratios. The transmission models have been renormalized to match the average $1-2~\mum$ transit depth, for the sake of comparing the spectral shapes. 
In the optical-IR ($0.3-2~\mum$), lab-grown tholins exhibit a relatively clear transmission (low $k$) window that shifts towards longer wavelengths as the C/O ratio increases. However, the transmission spectrum in this wavelength range appears smooth, because scattering by $0.1~\mum$ scale particles dominates.
At longer wavelenghts, hazes in a near-solar C/O atmosphere are predicted to exhibit a strong and narrow peak in the transmission spectrum around $5.8~\mum$. As the abundance of oxygen decreases, this feature broadens and shifts redder by approximately $0.3~\mum$. 
The relatively flat $6-8~\mum$ absorption profile of the tholin species examined here makes it so that molecular features dominate the observed spectral shape around $8~\mum$. 
Thus overall, transmission spectra captured across the infrared wavelength range of $3-10~\mum$ will be more suitable for identifying haze species and C/O ratios in \GJtwelve.

While Figure~\ref{fig:GJ1214} does not demonstrate a real fit to the \GJtwelve\ transit data, a few trends are visible.
The $0.5-2~\mum$ data are more consistent with the near-Solar and C/O$=1$ transmission models than those that employ haze opacities from tholins grown in Titan-like environments, where C/O$=\infty$. 
However, the Spitzer data are more consistent with our high metallicity $Z=1000$ calculation, while the high precision $1-2~\mum$ data are not (Figure~\ref{fig:GJ1214}, bottom). This is consistent with the findings of \citet{Lavvas2019}, who then require a haze formation yield $\sim 10-20\%$ in order to match the transmission wavelength at shorter wavelengths. The \citet{Kawashima2018} models used in this work are more consistent with $1\%$ haze formation efficiency scenarios from \citet{Lavvas2019}, with the caveat that average particle sizes from K19 are a factor of 3-10 larger. 
The JWST observations of \GJtwelve\ \citep{Greene2017, Bean2021} are likely to provide firmer insight for distinguishing among haze formation models and C/O ratios. 

As noted in Section~\ref{sec:simulations}, the expected molecular abundances in sub-Neptune atmospheres are not identical to the laboratory environment in most tholin experiments, which are designed to mimic Titan and early-Earth conditions. 
The presence of particular molecules, not just bulk C:N:O ratios, affects the optical properties of photochemical hazes, their production efficiency, and particle size distribution \citep{Horst2014, Brasse2015, Horst2017a, Ugelow2018}. Recent experiments that simulate photochemical conditions in hot Jupiters with $T > 1000$~K inject doubt that hazes can form from CO and H$_2$O in an H$_2$ dominated atmosphere \citep{Fleury2019, Fleury2020}. Our transmission spectrum models rely on the theoretical predictions of \citet{Kawashima2019b}, where hazes do form from HCN, C$_2$H$_2$, and CH$_4$ in an H$_2$ dominated atmosphere and temperatures spanning 500-1200~K. Other aerosol production models also predict that Jupiter-like planets can form hazes from CH$_4$ at $T_{\rm eq} < 950$~K \citep{Gao2020}. 
Despite these nuances, the optical properties of K84 are widely used as a template for chemical hazes in non-terrestrial environments. While the optical constants provided by this work are similarly imperfect for use with gas giants and sub-Neptunes, they provide a necessary advancement that is an improvement over the current practice.

Based on the laboratory data, we expect to be able to distinguish among hazes grown in different C/O ratio environments via strong C=O resonances observable around $6~\mum$, arising from the enhanced uptake of oxygen into the solid phase. 
If present, distinguishing among haze species in the atmosphere of a cool ($< 800$~K) planet with a H/He rich envelope is most plausible with the current generation of telescopes. We demonstrate that a model  \GJtwelve\ atmosphere that employs K84 optical constants to predict transmission spectra under the influence of haze obscuration could differ by 200-1500~ppm and, in the case of a solar C/O atmosphere, be underestimated by as much as 10\%.
If the atmosphere of \GJtwelve\ has a high mean molecular weight, represented by our Z=1000 simulation, $\sim 20$~ppm sensitivity is required to distinguish among haze species. 
This level of precision may be achievable with JWST for a select number of transiting sub-Neptunes. 
Based on the experimental setup, the optical properties for these lab-grown tholins are even more relevant for temperate terrestrial planets, which will only be accessible by future generations of ground and space-based telescopes. 
The optical constants and size-dependent cross-sections of the tholins used in this work are publicly available in several formats that can be adapted for use by other open source transmission modeling tools. A static version of eblur/newdust and the custom version of ExoTransmit, used to compute the cross-sections and transmission spectra, are also provided with this data release (doi:10.5281/zenodo.7500026).

\bibliography{references}{}
\bibliographystyle{aasjournal}

\end{document}